\documentclass[aps,pre,amsmath,amssymb,onecolumn]{revtex4}
\usepackage{graphicx}
\usepackage{subfigure}
\newcommand{\be}{\begin{equation}}\newcommand{\ee}{\end{equation}}
\newcommand{\ba}{\begin{array}{l}}\newcommand{\ea}{\end{array}}
\newcommand{\baa}{\begin{eqnarray}}\newcommand{\eaa}{\end{eqnarray}}
\newcommand{\lab}[1]{\label{#1}}\newcommand{\re}[1]{(\ref{#1})}
\newcommand{\ci}[1]{\cite{#1}}

\begin{document}

\title {Networks with point like nonlinearities}
\author{K.K. Sabirov$^{1}$, J.R. Yusupov$^{2}$, H. Susanto$^{3}$ and D.U. Matrasulov$^{2}$}

\affiliation{$^1$ Tashkent University of Information Technologies,
108 Amir Temur Str., 100200, Tashkent Uzbekistan\\
$^2$ Turin Polytechnic University in Tashkent, 17 Niyazov Str., 100095, Tashkent, Uzbekistan\\
$^3$Department of Mathematical Sciences, University of Essex, Colchester, CO4
3SQ, United Kingdom}

\begin{abstract}
We study static nonlinear waves in networks described by a nonlinear
Schr\"odinger equation with point-like nonlinearities on metric graphs.
Explicit solutions fulfilling vertex boundary conditions are obtained.
Spontaneous symmetry breaking caused by bifurcations is found.
\end{abstract}


\maketitle

\section{Introduction}

Modelling wave and particle transports in branched structures is an important
problem with applications in many subjects of contemporary physics, such as
optics, condensed matters, complex molecules, polymers and fluid dynamics.
Mathematical treatment of such problems is reduced to solving different partial
differential equations (PDEs) on so-called metric graphs. These are set of
one-dimensional bonds, with assigned lengths. The connection rule of the bonds
is called topology of the graph and is described in terms of the adjacency
matrix \cite{Uzy1,Exner15}. Linear and nonlinear wave equations on metric
graphs attracted much attention recently and they are becoming a hot topic
\cite{Hadi1}-\cite{Noja19}.

Solving wave equations on metric graphs requires imposing boundary conditions
at the branching points (graph vertices). In case of linear wave equations,
e.g., for linear Schr\"odinger equation the main requirement for such boundary
conditions is that they should keep self-adjointness of the problem \ci{Kost},
while for nonlinear PDEs one needs to use other fundamental conservation laws
(e.g., energy, norm, momentum, charge, etc) for obtaining vertex boundary
conditions \ci{Zarif,Our1,KarimNLDE}. Energy and norm conservation was used to
derive vertex boundary conditions for the nonlinear Schr\"odinger equation
(NLSE) on metric graphs in \ci{Zarif}, where exact solutions were obtained and
integrability of the problem was shown under certain constraints. In \ci{Our1}
a similar study was done for the sine-Gordon equation on metric graphs. Soliton
solutions  of a nonlinear Dirac equation on metric graphs have been obtained in
\ci{KarimNLDE}. Static solitons in networks were studied in the Refs.
\ci{Adami2011}-\ci{Adami16} by solving stationary nonlinear Schr\"odinger
equations on metric graphs. Other wave equations on metric graphs have been
studied recently in \ci{Karim2018}. Modelling nonlinear waves and solitons in
branched structures and networks provides powerful tool for tunable  wave,
particle, heat and energy transport in different practically important systems,
such as branched optical fibres, carbon nanotube networks, branched polymers
and low-dimensional functional materials.

Here we consider a Schr\"odinger equation with point-like nonlinearity on
metric graphs. NLSE with pointlike nonlinearity can be implemented in dual-core
fiber Bragg gratings as well as in the ordinary fibers \ci{Malomed3} and
Bose-Einstein condensates confined in double-well traps \ci{Malomed4,Malomed5}.
A similar problem on a line with double-delta type nonlinearity was considered
earlier in \ci{Malomed1,Malomed2}, where explicit solutions were derived. On
the basis of numerical analysis, it was shown that symmetric states are stable
up to a spontaneous symmetry breaking  bifurcation point. From a fundamental
viewpoint it would be interesting to see the difference between solutions of
the problem on the line and networks, as the topology of a network may cause
additional effects. Here, we use the methods of \ci{Malomed1} to obtain
explicit solutions of our problem. Degenerate spontaneous symmetry breaking
bifurcations are also obtained.

The paper is organized as follows. In the next section formulation of the
problem for metric star graph and the derivation of the vertex boundary
conditions are presented. In Section \ref{sec3} we obtain exact analytical
solutions of the problem and formulate constraints for integrability. Numerical
results and analysis of bifurcation is also presented in this section. Finally,
Section \ref{sec4} presents some concluding remarks.

\section{Vertex boundary conditions}
\label{sec2}

\begin{figure}[b!]
\centering
\includegraphics[width=100mm]{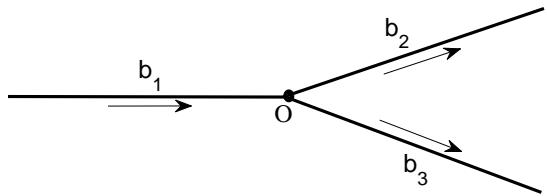}
\caption{Metric star graph} \label{pic1}
\end{figure}

Consider a  metric star graph consisting of three semi-infinite bonds,
$b_1\sim(-\infty;0),\,b_3\sim(0;+\infty),\,b_3\sim(0;+\infty)$ (see,
Fig.~\ref{pic1}).

On each bond of this graph the NLSE with variable nonlinearity coefficient,
$g_j(x)$ can be written as
\begin{equation}
i\frac{\partial\psi_j}{\partial t}=-\frac{1}{2}\frac{\partial^2\psi_j}{\partial
x^2} + g_j(x)|\psi_j|^2\psi_j,\label{eq1}
\end{equation}
where $j=1,2,3$ denotes the bond number.

To solve Eq.~\re{eq1}, one needs to impose vertex boundary conditions (VBC),
which can be derived, e.g., from norm and energy conservations laws. The norm
and energy are given respectively by
\begin{equation}
N=\underset{j=1}{\overset{3}{\sum}}\underset{b_j}{\int}|\psi_j(x)|^2dx,\label{eq3}
\end{equation}
and
\begin{equation}
H=\frac{1}{2}\underset{j=1}{\overset{3}{\sum}}\underset{b_j}{\int}\left(\left|\frac{\partial\psi_j}{\partial
x}\right|^2+g_j(x)|\psi_j|^4\right)dx.\label{eq4}
\end{equation}
From $\dot N = 0$ and $\dot H = 0$ and using
$\psi_1,\,\frac{\partial\psi_1}{\partial x}\to0$ as $x\to-\infty$ and
$\psi_{2,3},\,\frac{\partial\psi_{2,3}}{\partial x}\to0$ as $x\to+\infty$,  we
obtain the following vertex boundary conditions (at $x=0$)
\begin{eqnarray}
&{\bf Im}\left(\psi_1\frac{\partial\psi_1^*}{\partial x}\right)={\bf
Im}\left(\psi_2\frac{\partial\psi_2^*}{\partial x}\right)+{\bf
Im}\left(\psi_3\frac{\partial\psi_3^*}{\partial
x}\right),\label{eq5}\\
&{\bf Re}\left(\frac{\partial\psi_1}{\partial
t}\frac{\partial\psi_1^*}{\partial x}\right)={\bf
Re}\left(\frac{\partial\psi_2}{\partial t}\frac{\partial\psi_2^*}{\partial
x}\right)+{\bf Re}\left(\frac{\partial\psi_3}{\partial
t}\frac{\partial\psi_3^*}{\partial x}\right).\label{eq6}
\end{eqnarray}
Thus both, energy and current conservation give rise to nonlinear vertex
boundary conditions. However, the VBC given by Eqs.~\re{eq5} and \re{eq6} can
be fulfilled if the following two types of the linear relations at the vertices are imposed:\\
Type I:
\begin{eqnarray}
\left\{\begin{array}{cc}\alpha_1\psi_1|_{x=0}=\alpha_2\psi_2|_{x=0}=\alpha_3\psi_3|_{x=0},\\
\left.\frac{1}{\alpha_1}\frac{\partial\psi_1}{\partial
x}\right|_{x=0}=\left.\frac{1}{\alpha_2}\frac{\partial\psi_2}{\partial
x}\right|_{x=0}+\left.\frac{1}{\alpha_3}\frac{\partial\psi_3}{\partial
x}\right|_{x=0},\end{array}\right.\label{eq7}
\end{eqnarray}
and\\
Type II\\
\begin{eqnarray}
\left\{\begin{array}{cc}\frac{1}{\alpha_1}\psi_1|_{x=0}=\frac{1}{\alpha_2}\psi_2|_{x=2}+\frac{1}{\alpha_3}\psi_3|_{x=0},\\
\left.\alpha_1\frac{\partial\psi_1}{\partial
x}\right|_{x=0}=\left.\alpha_2\frac{\partial\psi_2}{\partial
x}\right|_{x=0}=\left.\alpha_3\frac{\partial\psi_3}{\partial
x}\right|_{x=0},\end{array}\right.\label{eq8}
\end{eqnarray}
where $\alpha_1, \alpha_2, \alpha_3$ are real constants which will be
determined below. In the following we will focus on VBC of type I, as it looks
more physical. In the next section we obtain exact analytical solutions of
Eq.~\re{eq1} for the VBCs given by Eq.~\re{eq7} and derive a constraint which
provides integrability of the problem.

\section{Exact solutions and bifurcations}
\label{sec3}

Consider a localised nonlinearity given by
$$g_j(x)=-\frac{\beta_j}{a\sqrt{\pi}}\left[e^{-\frac{(x+c)^2}{a^2}}+e^{-\frac{(x-c)^2}{a^2}}.\right]$$
For this specific form of $g_j(x)$, space and time variables in Eq.~(\ref{eq1})
can be separated and for $a\to0$ it can be reduced to the following form:
\begin{equation}
-\mu\phi_j+\frac{1}{2}\phi_j''+\beta_j\left(\delta(x+c)+\delta(x-c)\right)\phi_j^3=0.\label{eq18}
\end{equation}
The solution of Eq.~(\ref{eq18}) without the vertex boundary conditions can be
written as \cite{Malomed1}
\begin{align}
\phi_1(x)&=\left\{\begin{array}{ll}\frac{B_1}{\sqrt{\beta_1}}e^{\sqrt{2\mu}(x+c)},\quad
x<-c,\\\frac{A_{11}}{\sqrt{\beta_1}}e^{-\sqrt{2\mu}(x-c)}+\frac{A_{12}}{\sqrt{\beta_1}}e^{\sqrt{2\mu}(x+c)},\quad
0\geq x>-c,
\end{array} \right.\nonumber\\
\phi_j(x)&=\left\{\begin{array}{ll}\frac{B_{j1}}{\sqrt{\beta_j}}e^{-\sqrt{2\mu}(x-c)}+\frac{B_{j2}}{\sqrt{\beta_j}}e^{\sqrt{2\mu}(x+c)},\quad
0\leq x<c,\\\frac{A_{j}}{\sqrt{\beta_j}}e^{-\sqrt{2\mu}(x-c)},\quad
x>c.\end{array} \right.,j=2,3.\label{solution3}
\end{align}

Fulfilling the VBCs \re{eq7} by solutions \re{solution3} leads to
\begin{eqnarray}
\frac{\alpha_1}{\sqrt{\beta_1}}(A_{11}+A_{12})=\frac{\alpha_2}{\sqrt{\beta_2}}(B_{21}+B_{22})=\frac{\alpha_3}{\sqrt{\beta_3}}(B_{31}+B_{32}),\nonumber\\
\frac{1}{\alpha_1\sqrt{\beta_1}}(-A_{11}+A_{12})=\frac{1}{\alpha_2\sqrt{\beta_2}}(-B_{21}+B_{22})+\frac{1}{\alpha_3\sqrt{\beta_3}}(-B_{31}+B_{32}).\label{eq19}
\end{eqnarray}

Choosing  parameters $A$ and $B$ to fulfil the relations will yield
$$
B_{j1}=A_{11},\quad A_{12}=A\cdot A_{11},\quad B_{j2}=A\cdot B_{j1},\, A\neq
\pm 1 \;\;j=2,3.
$$
From the first equation of (\ref{eq19}) we get
$$
\frac{\alpha_1}{\alpha_{2,3}}=\frac{\sqrt{\beta_1}}{\sqrt{\beta_{2,3}}},
$$
$$
\frac{1}{\alpha_1\sqrt{\beta_1}}=\frac{1}{\alpha_2\sqrt{\beta_2}}+\frac{1}{\alpha_3\sqrt{\beta_3}}.
$$
These equations lead to the constraint  given by \be
\frac1\beta_1=\frac1\beta_2+\frac1\beta_3. \lab{sr01}\ee

Furthermore, from the continuity of the solution $\phi_j(x)$ we have
\begin{eqnarray}
B_1=A_{11}(e^{2\sqrt{2\mu}c}+A),\, A_{2,3}=A_{11}(A\cdot
e^{2\sqrt{2\mu}c}+1).\nonumber
\end{eqnarray}

For the jump $\Delta(\phi_1')|_{x=-c}=-2\beta_1\left(\phi_1|_{x=-c}\right)^3$
we can find
\begin{equation}
A_{11}=\pm\sqrt{\frac{\sqrt{2\mu}e^{2\sqrt{2\mu}c}}{\left(e^{2\sqrt{2\mu}c}+A\right)^3}}.\label{coef1}
\end{equation}
For the jump
$\Delta(\phi_j')|_{x=c}=-2\beta_j\left(\phi_j|_{x=c}\right)^3,\,j=2,3$ we can
find
\begin{equation}
A_{11}=\pm\sqrt{\frac{\sqrt{2\mu}Ae^{2\sqrt{2\mu}c}}{\left(Ae^{2\sqrt{2\mu}c}+1\right)^3}}.\label{coef2}
\end{equation}
Equating (\ref{coef1}) and (\ref{coef2}) we have
\begin{equation}
A=\frac{e^{6\sqrt{2\mu}c}-3e^{2\sqrt{2\mu}c}\pm\sqrt{\left(3e^{2\sqrt{2\mu}c}-e^{6\sqrt{2\mu}c}\right)^2-4}}{2},\label{eq20}
\end{equation}
where $\mu\geq\frac{\ln^22}{8c^2}$.

Similarly as the above, one can obtain solutions for the cases of one- and two
nonlinear bonds. For the graph with two nonlinear bonds we have the following
stationary NLSE on each bond $b_j$
($b_1\sim(-\infty;0],\,b_{2,3}\sim[0;+\infty)$) of the star graph
\begin{eqnarray}
-\mu\phi_1+\frac{1}{2}\phi_1''+\beta_1\delta(x+c)\phi_1^3=0,\nonumber\\
-\mu\phi_2+\frac{1}{2}\phi_2''+\beta_2\delta(x-c)\phi_2^3=0,\nonumber\\
-\mu\phi_3+\frac{1}{2}\phi_3''+\beta_3\delta(x+c)\phi_3^3=0,\,c>0.\label{twoeq1}
\end{eqnarray}
Under the constraints given by Eq.~\re{sr01} the solutions fulfilling the
boundary conditions can be written as
\begin{eqnarray}
\phi_1(x)=\left\{\begin{array}{cc}\frac{A}{\sqrt{\beta_1}}e^{\sqrt{2\mu}(x+c)},\text{at}\,x<-c,\\\frac{A}{\sqrt{\beta_1}}e^{-\sqrt{2\mu}(x+c)},\text{at}\,0\geq
x<-c,\end{array}\right.\nonumber\\
\phi_2(x)=\left\{\begin{array}{cc}\frac{A}{\sqrt{\beta_2}}e^{\sqrt{2\mu}(x-c)},\text{at}\,0\leq
x<c,\\\frac{A}{\sqrt{\beta_2}}e^{-\sqrt{2\mu}(x-c)},\text{at}\,x>c,\end{array}\right.\nonumber\\
\phi_3(x)=\frac{A}{\sqrt{\beta_3}}e^{-\sqrt{2\mu}(x+c)},\text{at}\,x\geq0.\label{twoeq2}
\end{eqnarray}

\begin{figure}[t!]
    \centering
    \subfigure[]{\includegraphics[scale=0.39,clip=]{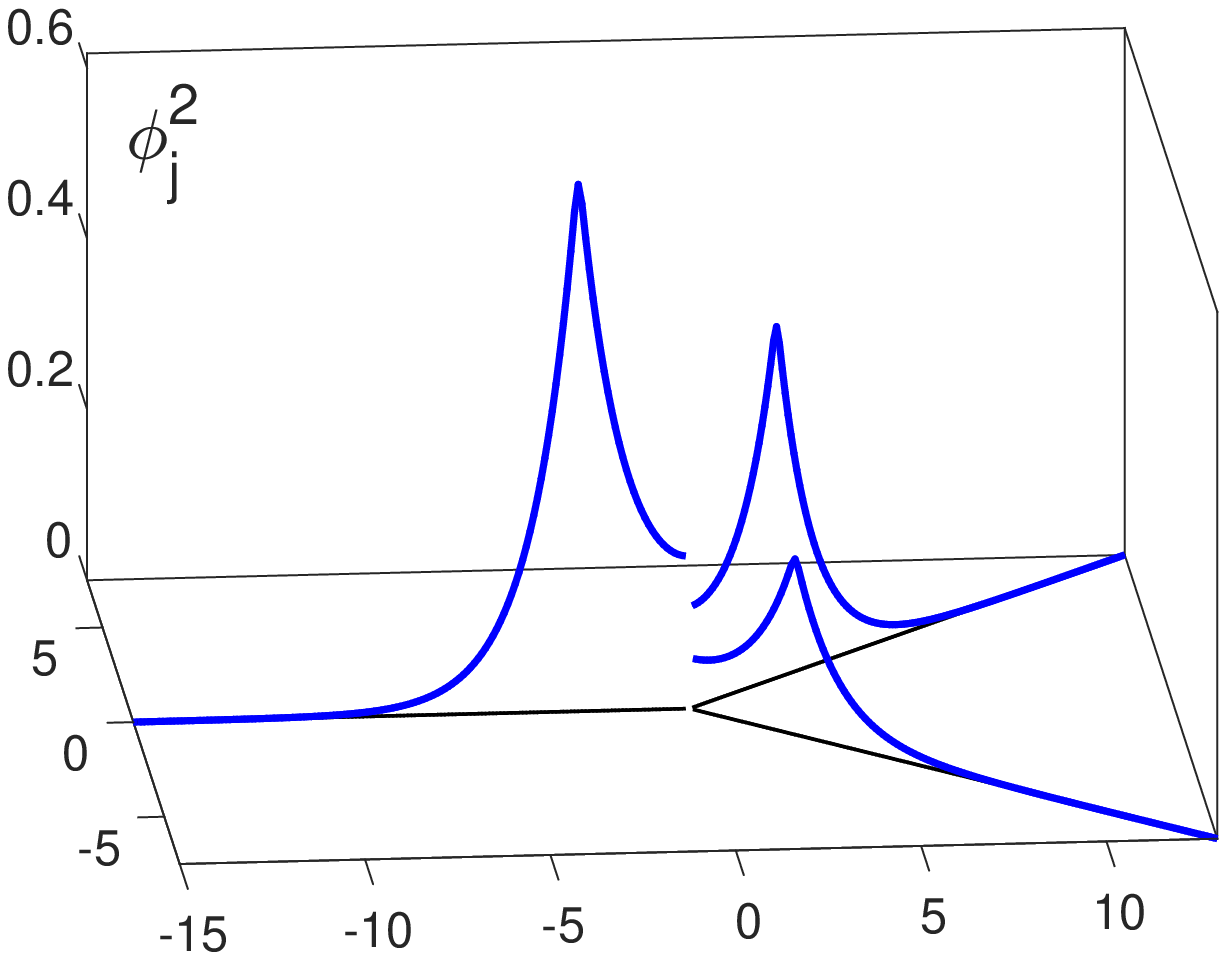}\label{fig2a}}
    \subfigure[]{\includegraphics[scale=0.39,clip=]{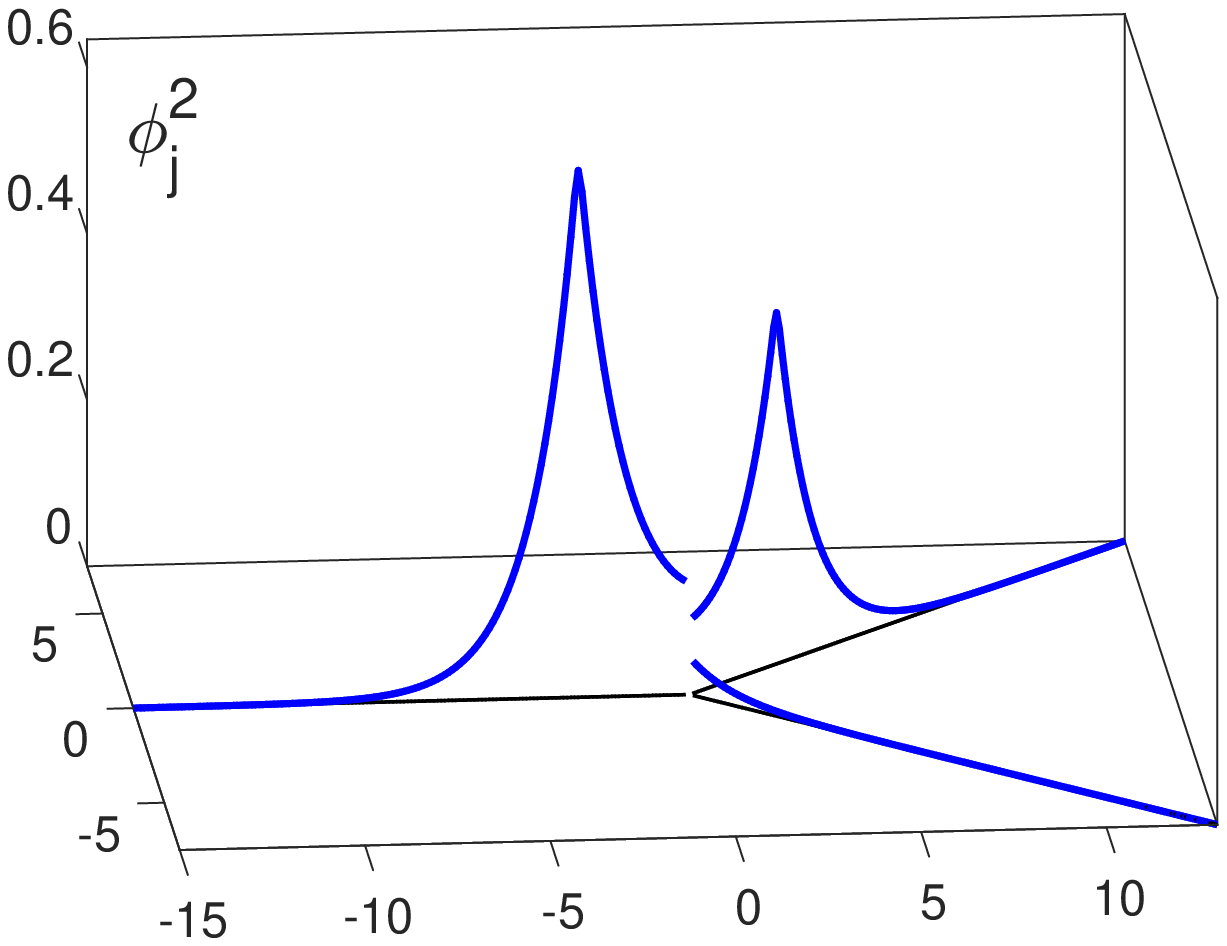}\label{fig2b}}
    \caption{Two possible solutions are plotted. Here, $\beta_1=2/3$, $\beta_2=1$,  and $\beta_3=2$ satisfying the condition for conserved norm and energy, and $\mu=0.1$. The solution in panel (a) is denoted by configuration $[1,1,1]$ and in panel (b) by $[1,1,0]$.}
    \label{fig2}
\end{figure}

For a single nonlinear bond we have the solution
\begin{eqnarray}
\phi_1(x)=\left\{\begin{array}{ll}\frac{B}{\sqrt{\beta_1}}e^{\sqrt{2\mu}(x+c)}\,{\rm
at}\,x<-c,\\\frac{A}{\sqrt{\beta_1}}e^{-\sqrt{2\mu}(x+c)}\,{\rm at}\,0\geq
x>-c,\end{array}\right.\nonumber\\
\phi_{2,3}(x)=\frac{A}{\sqrt{\beta_{2,3}}}e^{-\sqrt{2\mu}(x+c)}\,{\rm
at}\,x\geq0.\label{eq12}
\end{eqnarray}
From the  continuity of  $\phi_1(x)$ at $x=-c$ we have
$$
B=A.
$$
One can find  $A$ from the expression for the norm given by Eq.~\re{eq3} :
$$
A=\sqrt{\frac{2N\sqrt{2\mu}}{\left(\frac{1}{\beta_2}+\frac{1}{\beta_3}-\frac{1}{\beta_1}\right)e^{-2\sqrt{2\mu}c}+\frac{2}{\beta_1}}}.
$$

The above analytical solution is obtained under the assumption that the
constraint \re{sr01} is satisfied. In the following, we solve Eq.~\re{eq18}
numerically both for the case when the constraint in Eq.~\re{sr01} is fulfilled
and broken. The point nonlinearity in Eq.~\re{eq18} is represented by the
Gaussian function with $c=3$ and $a=0.1$. In the following, we only limit
ourselves with positive solutions.

In Fig.~\ref{fig2}, we plot two possible solutions for the case when the sum
rule given by Eq.~\re{sr01} is fulfilled. The first panel shows a solution when
all the bonds are excited by the delta nonlinearity, while in the second one
(b), only the first and the second bonds have point-like excitations. Using the
configuration in the limit $\mu\to\infty$ as our code, we represent solutions
in panels (a) and (b) as $[1,1,1]$ and $[1,1,0]$, respectively.

\begin{figure}[t!]
    \centering
    \subfigure[]{\includegraphics[scale=0.39,clip=]{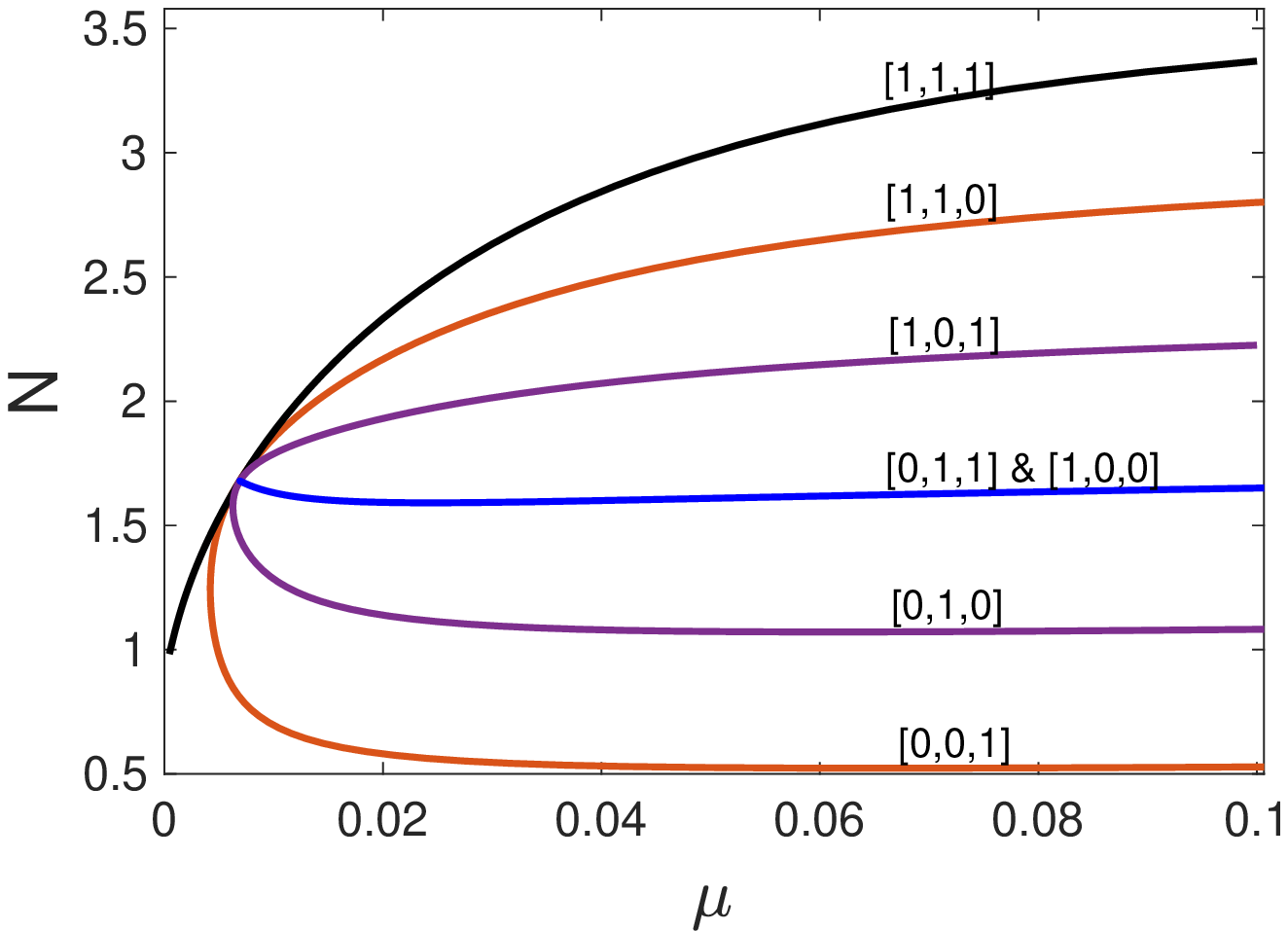}\label{fig3a}}
    \subfigure[]{\includegraphics[scale=0.39,clip=]{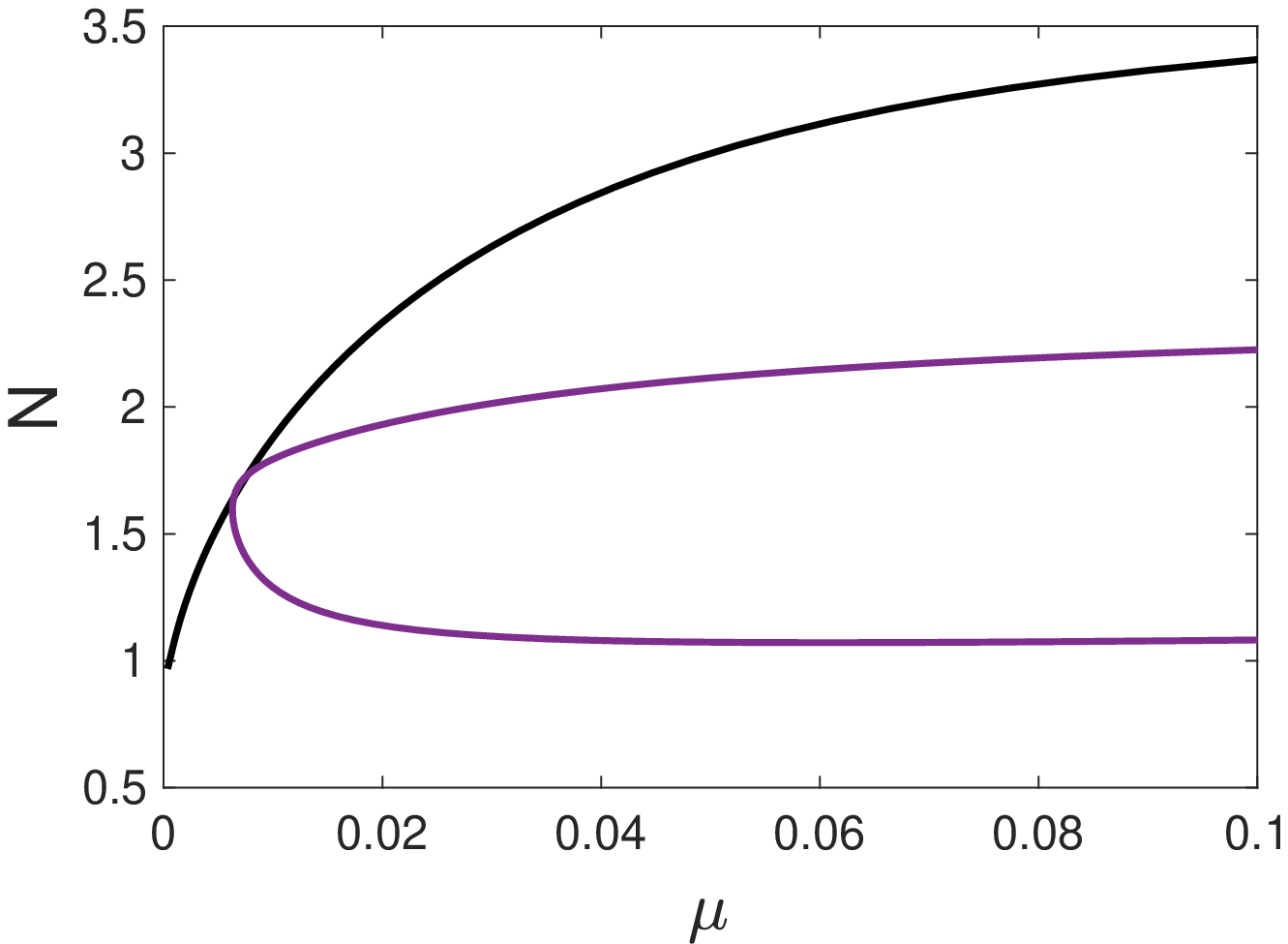}\label{fig3b}}
    \caption{(a) Bifurcation diagram of solutions in Fig.\ \ref{fig2}. (b) The same diagram, but for $\beta_j=1$, $j=1,2,3.$}
    \label{fig3}
\end{figure}

In Fig.~\ref{fig3a}, we present bifurcation diagrams of the solutions in
Fig.~\ref{fig2}. We obtain that the two configurations in Fig.~\ref{fig2} are
connected with each other, with $[1,1,1]$ as the main branch and $[1,1,0]$ as a
bifurcating solution through a pitchfork bifurcation with the configuration
$[0,0,1]$. We therefore observe a spontaneous symmetry breaking bifurcation. It
is particularly interesting to note that the bifurcation is quite degenerate in
the sense that we obtain both a subcritical as well as a supercritical
bifurcation emerging from the same point. We also obtain several other
solutions bifurcating from the same bifurcation point, which are all indicated
in Fig.~\ref{fig3a}.

We have considered a different case when all the nonlinearity coefficients are
the same, i.e., without loss of nonlinearity $\beta_j=1$. In this case, the
condition \re{sr01} is not satisfied. We plot the bifurcation diagram of the
positive solutions in Fig.~\ref{fig3b}, where now we obtain that all the
asymmetric solutions merge into two branches only, which bifurcate from the
same point.

\section{Conclusions}
\label{sec4}

In this paper we obtained and analyzed  exact solutions  of NLSE on metric
graphs with varying nonlinearity that has the form of a delta-well. Exact
analytical solutions of the problem were obtained for different cases of point
excitations, determined by the presence of a delta-well on different bonds. The
constraint providing existence of such analytical solutions are derived in the
form of simple sum, rule written in terms of the bond nonlinearity
coefficients. Numerical solutions of the problem are also obtained both for
integrable and non-integrable cases. Bifurcations of the solutions are studied
in terms of chemical potential, using the numerical solutions $\mu$ using the
numerical solutions. The model considered in this paper is relevant for
different practically important problems such as BEC in branched traps, Bragg
gratings in branched fibers, etc. Extension of the treatment to other graphs
topologies is rather straightforward, provided the graphs contains arbitrary
subgraph, which is connected to three or more outgoing semi-infinite bonds.

\section{Acknowledgements}
This work is partially supported by a grant of the Ministry of Innovation
Development of Uzbekistan (Ref. No. BF-2-022).

\end{document}